%% file: submit5.tex
\documentclass[
reprint,
nofootinbib,
amsmath,amssymb,amsbsy,
prd,
onecolumn,
superscriptaddress,
longbibliography,
preprintnumbers]{revtex4-1}
\usepackage{xcolor}
 \newcommand{\bcf}{$B \chi F$}

\input{preamble}

\begin{document}

\title{Regulated chiral gauge theory and the strong CP problem}

\author{David B. Kaplan}
\email{dbkaplan@uw.edu}
\affiliation{Institute for Nuclear Theory, Box 351550, Seattle, Washington 98195-1550}

\author{Srimoyee Sen,}
\email{srimoyee08@gmail.com}
\affiliation{Department of Physics and Astronomy,  Iowa State University, Ames, Iowa 50011}

\date{\today}
\preprint{INT-PUB-24-040}

\begin{abstract}
{Four-dimensional chiral gauge theory can be formulated as the boundary theory on a five-dimensional manifold in a manner that may be realized on a finite lattice. There are interesting features of these theories which defy a purely four-dimensional conception of universality. We find that QCD when embedded in a chiral gauge theory (the Standard Model) and regulated this way can simultaneously avoid both the $U(1)_A$ problem and the strong $CP$ problem, with a central role played by fermion zeromodes localized far away in the fifth dimension. In this way it differs from conventional lattice QCD formulated as a stand-alone theory, universality being violated by inaccessible light modes in the five-dimensional bulk. Our analysis  builds on recent work by others that highlights the role of global $U(1)$ symmetries in five dimensional formulations of four-dimensional chiral gauge theories, and the generic appearance  of fermion zeromodes in the bulk.}

\end{abstract}
\maketitle

%%%%%%%%%%%%%%%%%

\section{Introduction}

%%%%%%%%%%%%%%%%%

Despite the Standard Model enjoying tremendous success, {a nonperturbative regulator   for  the theory   has never been demonstrated to exist.} This is a disturbing situation for the theory we believe describes the non-gravitational interactions of elementary particles in the {real} world.    While accelerator phenomenology does not currently require high order perturbative computations that are sensitive to the chirality of fermions, there are nonperturbative computations that would be interesting to perform, such as the baryon number violation rate of the Standard Model in the early universe.   Specific computations aside, however, we cannot claim to thoroughly understand the theory without understanding how {\it in principle}  to compute its implications  in a controlled manner to arbitrary precision.   We are aware that there are strong constraints on the fermion content arising from requiring gauge anomaly cancellation, but we cannot know with certainty that these are the only constraints that must be met without having control over the UV properties of the theory. Attempting to settle this issue is our primary motivation  for seeking a nonperturbative regulator for chiral gauge theories.    

Over the past half century there have been  very many  attempts to devise such a regulator for chiral gauge theories. By the early 1980s it was already clear that the problem was not merely one of algorithm development, but that  a fundamental obstacle existed to representing fermions on the lattice with exact chiral symmetry \cite{karsten1981lattice,Nielsen:1980rz}, and subsequent attempts have focused on circumventing the no-go theorem. We will not attempt to catalog those attempts here.

This article concerns itself with the specific proposal for a lattice regulator for chiral gauge theories recently proposed in Ref.~\cite{kaplan2024chiral}. Wilson fermions are simulated on a discretized five-dimensional Euclidian manifold $X$ with a single connected four-dimensional boundary $\delta X$ ({\it e.g.} $X$ could be a filled 4-torus with a 4-torus boundary for $\delta X$, or $X$ could be a 5-ball with $\delta X$ being the 4-sphere boundary \cite{clancy2025chiral}).   Open boundary conditions are imposed on the fermions and the ratio $M/r$ is chosen to lie in a range corresponding to a nontrivial topological phase \cite{Kaplan:1992bt,Jansen:1992tw,Golterman:1992ub}, where $M$ is the mass and $r$ is the coefficient of the Wilson term.  Such a configuration has been shown to support Weyl edge states without mirror fermions, while all the bulk states are gapped at the cutoff  \cite{kaplan2024weyl}; see \cite{aoki2022curved,aoki2023curved,aoki2024,aoki2024study} for related work.  To ensure that the boundary theory looks like a conventional four-dimensional gauge theory, the proposal is to  integrate over the boundary gauge fields $A_\mu$ in the path integral, while for each configuration, the bulk gauge fields $B_k$ are constrained by the classical 5d lattice Yang-Mills equations of motion, subject to the 4d boundary values \cite{grabowska2016nonperturbative,kaplan2024chiral}\footnote{We use Latin indices $k=1,\ldots,5$ to designate coordinates on the 5d manifold $X$, and Greek indices $\mu=1,\ldots,4$ for coordinates on the boundary $\delta X$.}. Thus the gauge field that  the bulk fermions see is a nonlocal functional $B_k[A_\mu]$ of the boundary fields. The integration over boundary fields is then weighted by $\Delta[B] \exp(-S_\text{YM}[A])$, where $S_\text{YM}$ is the conventional 4d  Wilson plaquette action and $\Delta$ is the five-dimensional  fermion determinant.
  We will refer to this proposal as the ``boundary chiral fermion'' {proposal, \bcf~ for short.
  
 With the bulk gapped, $\ln\Delta[B]$ can be expanded in local gauge invariant operators.  The leading contribution to the imaginary part is the  Chern-Simons operator which gives rise to bulk currents that exactly compensate for anomalous charge violation on the boundary.  This phenomenon,  now called ``anomaly inflow'',   was discovered for  continuum theories by Callan and Harvey \cite{Callan:1984sa} (see \cite{witten_anomaly_2020} for a more recent discussion) and first demonstrated for a  lattice gauge theory in Ref.~\cite{jansen1992chiral}.   It is precisely when gauge anomalies cancel in the boundary theory that the coefficient of the Chern-Simons operator vanishes and one can plausibly obtain a local gauge theory on the 4d boundary in the continuum limit.  \bcf  ~thereby provides a simple answer to the obvious question of what would go wrong if gauge anomalies in the boundary theory did $not$ cancel: it could never  have a continuum limit {resembling} a local 4d theory.

Recently an  {interesting} objection was made to the \bcf ~proposal by Golterman and Shamir \cite{golterman2024conserved} (GS) on the grounds that it predicts an exact global $U(1)$ symmetry that does not exist in the target  {boundary} theory. A related phenomenon is that fermion zeromodes will appear in the bulk  in the presence of nontrivial gauge topology; both phenomena are familiar from  an earlier incarnation of the model \cite{grabowska2016nonperturbative,hamada2017axial}, and bulk zeromodes were both observed numerically for fermions on the boundary of a 3-ball in Refs.~\cite{aoki2023curved,aoki2024} by Aoki et al., and were  discussed in Ref.~\cite{golterman2024conserved}.  This phenomenon is mandated by the fact that the Dirac operator must have zero index on a $d$-dimensional manifold which is the boundary of $(d+1)$-dimensional manifold, after summing over all disconnected pieces.  In Ref.~\cite{aoki2023magnetic} it is argued that at least on the lattice, bulk gauge singularities essentially create a new boundary for the mirror zeromodes to bind to.  These bulk zeromodes will appear in the resultant 't Hooft operators \cite{grabowska2016nonperturbative},  {and render them invariant under} $U(1)$ symmetries that should be anomalous in the purely 4d target theory \cite{golterman2024conserved}.

GS use a particularly simple model to illustrate the $U(1)$ problem, consisting of two bulk Dirac fields with opposite sign mass, giving rise to a massless Dirac fermion on the boundary.  By introducing an $SU(3)$ gauge symmetry, the boundary theory would be expected to behave like QCD with one massless flavor in the \bcf ~proposal,  {exhibiting} chiral symmetry breaking, a massive $\eta'$ meson, and no Nambu-Goldstone bosons.  However, the  {5d} theory has two exactly conserved 5d  $U(1)$ currents, corresponding to the fermion currents of the two bulk Dirac fermions.   Out of these GS construct two exactly conserved and gauge invariant 4d currents by integrating over the extra dimension. The charges constructed from these currents act on the boundary fermions fields as the usual vector and axial charges of the boundary QCD theory, and by means of a conventional Ward-Takahashi (WT) identity they  {argue} that either the boundary theory  spectrum contains a Nambu-Goldstone boson, or else the quarks do not condense, neither option being consistent with a conventional 4d theory of QCD with one flavor of massless quark.

This is a  {puzzling} phenomenon, since the existence of an exact global $U(1)$ symmetry for each bulk fermion field  {in the 5d theory} is the cornerstone of the anomaly inflow phenomenon, whereby the Chern-Simons current flowing onto the boundary  {is guaranteed to match} 
anomalous symmetry violation in the boundary theory, and it is precisely this anomalous symmetry violation that led to 't Hooft's solution of the conventional $U(1)_A$ problem  {in 4d} \cite{t1976symmetry}.    

In this paper we examine the issue in detail and conclude that there are indeed two exactly conserved $U(1)$ symmetries, but that  nevertheless, the boundary theory  {in the \bcf~ construction} should behave as ordinary QCD with a massive $\eta'$ meson.  {Furthermore,  the theory can be constructed to have no strong $CP$ problem, with   the phase of the quark mass matrix being unphysical.}  The way this can occur is similar to  {how} the massless up quark solution to the strong $CP$ problem  {works}, where the  {zeromodes of  the massless up quark quench the contributions of gauge field configurations with nontrivial topology to the partition function, rendering the vacuum $CP$-invariant} \cite{banks1994missing}.   A massless up quark is ruled out phenomenologically because it leads to a hadron spectrum inconsistent with what is observed.  In the \bcf~model, constructed such that quark masses are boundary operators,  {there exists an exact bulk fermion zeromode localized far off in the fifth dimension that similarly enforces trivial topology on the vacuum state; however, since the bulk fermions do not participate in the hadron spectrum of the boundary theory,  this solution to the strong $CP$ problem is consistent with the expected phenomenology of $N_f=1$ QCD}.

In this paper  {we carefully reexamine} the $U(1)$ problem in the \bcf ~proposal and its connection to anomaly-inflow and $CP$-violation,  {as well as the apparent paradox of having an exact $U(1)_A$ symmetry which is spontaneously broken, without having a massless $\eta'$}.  We find this most simply  done by means of a  sequence of effective field theories:  first with the bulk fermions integrated out of the theory, which is subsequently  matched onto a chiral Lagrangian for the boundary theory after integration over the boundary  gauge fields. Since the existence of the unwanted symmetry has nothing to do with  {constructing the theory on a lattice}, we  {primarily} consider the theory in the (Euclidian) continuum,  {occasionally pointing out issues peculiar to a lattice theory}.

%%%%%%%%%%%%%%%%%

\section{The $\mathbf{U(1)_A}$ problem in effective theory after integrating out bulk fermions}
\label{U1A}

%%%%%%%%%%%%%%%%%%

Our starting point is the 5d model considered by Golterman and Shamir
(GS), two flavors of Dirac fermion $\psi^\pm$  { with large masses equal to $\pm m$ respectively, with $m>0$, coupled to background 5d $SU(3)$ gauge fields.} On the lattice, the bulk fermion mass  $m$ is is equated with the UV cutoff,  the inverse lattice spacing $1/a$.  We will take the manifold to be a solid 4-torus, $ (S^1)^3\times D^2$, parametrized by coordinates  {$\{x,y,z,\theta, r\}$, with $\{\theta,r\}$ being coordinates for the disk  {where $0\le r\le R$,} and $\{x,y,z\}$  {are} periodically identified.  The 4d boundary  {at $r=R$} is the 4-torus, parametrized by $\{x,y,z,\tau\}$, where $\tau = R\theta$.}  
The Lagrangian is taken to be
\beq
\CL = \bar\psi^+ \left(\slashed{D} + m\right)\psi^+ + \bar\psi^- \left(\slashed{D} - m\right)\psi^- +  \CJ_{k}\left( \bar\psi^+ \gamma_{k} \psi^+- \bar\psi^-\gamma_{k} \psi^-\right)\ ,
\eqn{eft}
\eeq
where $D=(\partial+ i B)$ is the $SU(N_c)$ covariant derivative with bulk gauge fields $B_k = B_k^a T_a$, the $SU(N_c)$ generators normalized as  $\Tr T_a T_b=\half\delta_{ab}$, with $k=1,\ldots,5$ being the bulk spacetime coordinate index.  The  fermions are subject to the boundary  conditions  $(1\mp \gamma_r)\psi^\pm\vert_{r=R} = 0$   with $\gamma_r = \hat r\cdot\vec\gamma$ \footnote{ This theory may be equivalently   defined with $\psi^\pm$ on  $R^5$ (vanishing at infinity) with fermion masses $\mp M\to \mp \infty$ respectively for $r>R$.}.   {Note that this theory manifestly respects an exact global $U(1)\times U(1)$ symmetry, corresponding to independent phase redefinitions of $\psi^\pm$.}     This theory has an exactly massless Dirac fermion confined to the boundary at $r=R$  \cite{kaplan2024chiral}\footnote{The formulation on the lattice is slightly different \cite{kaplan2024weyl}: one can work with open boundary conditions while the operators $(\slashed{D}\pm m)$ are replaced by $D_w^\pm = \slashed{D} \pm (1 + \half D^2)$, where here $\slashed{D}$ and $D^2$ are the conventional discretized version of the covariant Dirac operator and Laplacian respectively \cite{kaplan2024weyl,Golterman:1992ub}.}. 
   {We have included in \eq{eft} a} gauge-singlet vector field $\CJ_{k}$,  a source for the  { linear combination of the two $U(1)$ bulk currents that generates the $U(1)_A$ transformation of the boundary fermion.}  

Assuming that all the bulk states are fully gapped -- an assumption we will reexamine below --  {we can integrate out the bulk fermions and arrive at the effective theory for the boundary modes, described by the path integral}
\beq
 {\int dq d\bar q \, \Delta(B) e^{-S_{4d}(q,\bar q,A)}\ ,}
\eeq
 {where $q$ is the boundary Dirac fermion, $A_\mu$ with $\mu=1,\ldots,4$ is the boundary value of the 4-vector part of the 5d gauge field $B_k$, and $\Delta(B)$ is the fermion determinant obtained by integrating out the bulk fermions.  Since the bulk fermions are assumed to be gapped, we can expand $\ln \Delta$ in terms of local operators involving the 5d gauge field and its derivatives, integrated over the bulk coordinates.  These operators will in general include the Yang-Mills kinetic term, a Chern-Simons term, and higher derivative operators. }

 The lattice version of this theory is well-defined, but not so the continuum version.  The existence of a  theory of edge states requires that the phase of the fermion determinant have a topological origin. As was shown in \cite{Golterman:1992ub}, the dispersion relation for a massive fermion in $d$ Euclidian dimensions defines a map from momentum space to a $d$-sphere, and the one-loop graph that determines the coefficient of the Chern-Simons term is computing the winding number of that map -- which is only well defined when   momentum space is compact.  On a hypercubic lattice,  momentum space is a torus.  In the continuum, however, to render momentum space compact one needs a regulator.  A Pauli-Villars field, for example, turns momentum space into a sphere and is an acceptable solution.  If one sets the mass of the Pauli-Villars field to equal the bulk fermion mass in magnitude but with opposite sign  while subject to the same boundary condition (as in Ref.~\cite{witten_anomaly_2020}), then the result is to replace the fermion determinant by}
\beq
 {\Delta(B) \to \Delta(B)_\text{reg}\ ,\qquad \Delta(B)_\text{reg}\simeq \frac{\Delta(B)}{\Delta^*(B)}\ ,}
\eeq
 {which doubles its phase and cancels its modulus everywhere except for at the boundary.  Nonperturbatively the phase is given by the $\eta$-invariant of the bulk Dirac operator, which in a perturbative expansion reduces to the Chern-Simons operator.}

 {The story on the lattice is more complicated: because the lattice theory has a compact momentum space, the phase of the fermion determinant is well defined; however, the lattice regulator does not cancel the modulus of the fermion determinant.  To achieve that, one has to introduce a Pauli-Villars field solely for that purpose. This is needed to ensure that the fermion determinant does not contribute a radiatively generated Yang-Mills action for the gauge field in the bulk, which would have to be fine-tuned away to avoid spoiling the 4d interpretation of the boundary gauge field dynamics.  This lattice Pauli-Villars field operates very differently than the one introduced in the continuum, precisely because it does not affect the phase of the fermion determinant. We discuss in the appendix how that can be realized, and assume that on the lattice theory as in the continuum, the fermion determinant included in the path integral is a pure phase.}

With only the phase of the fermion determinant being of interest, which is parity-odd, we can follow the logic of Callan and Harvey, who discovered ``anomaly in-flow'' in Ref.~\cite{Callan:1984sa}, and focus on the low energy effective theory with the bulk fermions integrated out.  This effective theory consists of the light boundary modes, as well as gauge field operators in the bulk.  The lowest dimension bulk operators will be the Yang-Mills and the Chern-Simons operators. Following Callan and Harvey, here we focus on the  Chern-Simons operators which contribute to the phase of the fermion determinant; below we will discuss the Yang-Mills operators which do not contribute to the phase, and will ignore higher dimension operators; the latter will be unimportant in the continuum theory at low energy, while in a lattice implementation they will be subsumed into the perennial question of how to tame rough gauge fields by approximating a more perfect gauge action.  The Chern-Simons operators  involve both the gauge field $B_k$, which couples identically to both $\psi^+$ and $\psi^-$ in \eq{eft}, and  the source $\CJ_k$, which couples to the two fermions with opposite sign. The contributions of the bulk fermions to the Chern-Simons depend on  both the relative sign of the fermion coupling to the gauge field and source, and to the sign of the fermion mass; in addition, for the continuum calculation, we include a Pauli-Villars regulator for both $\psi^\pm$  \eq{Left}\footnote{To obtain the correct coefficient $\kappa$ for the Chern-Simons operator in the continuum theory one must include the contribution of a regulator for each Dirac fermion of the 5d theory, such as a Pauli-Villars field; Callan and Harvey \cite{Callan:1984sa} did not include a regulator in their calculation  and for this reason the coefficient we find for the Chern-Simons term differs from theirs  by a factor of 2.}.  The result for the effective theory is\footnote{Here we repeat the logic of the Callan-Harvey computation by treating the fermion mass as spatially constant in the loop diagram, and then substituting the step function form in the coefficient of the Chern-Simons operator.  This is incorrect near the boundary, where ``near'' is to be compared to the bulk gap scale, which is the radial extent of the boundary mode wave functions.  The present form is therefore accurate in the limit that one considers wavelengths infinitely long compared to the gap scale.  In using $\CL_\text{eft}$ the only assumption that must be made to correctly account for fermion number conservation in the $5d$-theory and anomaly in-flow is that the $r$-derivative of $\theta(r-R)$ is twice the delta function $\delta(r-R)$ multiplying the boundary theory, with $\theta(0)=0$. A regulated version of these functions discussed below is used for practical computations. }

 %{\textcolor{red}{The low energy theory content of $\ln\Delta$ includes the  boundary fermions coupled to gauge fields and a series of gauge invariant operators constructed out of the gauge fields only that arise out of integrating out the bulk fermions. The leading nontrivial operators among these are  Chern-Simons contributions. Thus, we can write a long wavelength action given by} }

\beq
\CL_\text{eft} 
%&=&\delta(R-r)\left[\bar q \gamma^\mu D_\mu  q + \CJ_\mu \bar q \gamma^\mu \gamma_5  q\right]+ \kappa \theta(R-r)  \left(\CO_\text{CS}[V^+] -\CO_\text{CS}[V^-]\right) \cr  &&\cr
&=&
\delta(R-r)\left[\bar q \gamma^\mu D_\mu  q + \CJ_\mu \bar q \gamma^\mu \gamma_5  q\right] + 2 \kappa\theta(R-r)  \left(3 \CO_{\CJ F F}
 +  \CO_{\CJ \CJ \CJ}\right)\ ,\qquad \kappa = \frac{i}{24\pi^2} \ ,
\eqn{Left}\eeq
where the  the index $\mu$ refers to the 4d boundary coordinates $x_\mu =\{\bfx,R\theta\}$, and $\CO_{\CJ \CJ \CJ}$, $\CO_{\CJ F F}$ are defined by
\beq
\CO_{\CJ \CJ \CJ} 
{\bf d\boldsymbol{ \omega}}_5 &=& \Tr\left[d{\boldsymbol \CJ}\wedge d{\boldsymbol \CJ}\wedge {\boldsymbol \CJ} +\frac{3}{2}d{\boldsymbol \CJ}\wedge {\boldsymbol \CJ}\wedge {\boldsymbol \CJ}\wedge {\boldsymbol \CJ}+\frac{3}{5} {\boldsymbol \CJ}\wedge {\boldsymbol \CJ}\wedge {\boldsymbol \CJ}\wedge {\boldsymbol \CJ}\wedge {\boldsymbol \CJ} \right]\ ,\cr &&\cr
\CO_{\CJ F F} \,{\bf d\boldsymbol{ \omega}}_5 &=&\Tr {\boldsymbol \CJ}\wedge {\mathbf F}\wedge {\mathbf F}\ ,
\eqn{COs}
\eeq
where ${\bf d\boldsymbol{ \omega}}_5 = {\bf dx}^1\wedge\ldots\wedge {\bf dx}^5$, 
the 1-forms {${\mathbf B}$} and ${\boldsymbol {\CJ}}$ are defined as
$
 {\mathbf B} = i T_a B_k {\bf dx}^k$ ,
 ${\boldsymbol {\CJ}} = i   \CJ_k {\bf dx}^k\ $
with  $\CJ_k$  proportional to the unit matrix in color space, and 
${\mathbf F} = d{\mathbf B}+{\mathbf B}\wedge {\mathbf B}$.
%$\CO_\text{CS}[B]$ is the coefficient of the 5d Chern-Simons form
%\beq
%\CO_\text{CS}[B] \,{\bf d\boldsymbol{ \omega}}_5 &=& \Tr\left[d{\mathbf B}\wedge d{\mathbf B}\wedge {\mathbf B} +\frac{3}{2}d{\mathbf B}\wedge {\mathbf B}\wedge {\mathbf B}\wedge {\mathbf B}+\frac{3}{5} {\mathbf B}\wedge {\mathbf B}\wedge {\mathbf B}\wedge {\mathbf B}\wedge {\mathbf B} \right] \ ,
%\eeq
%where ${\bf d\boldsymbol{ \omega}}_5 = {\bf dx}^1\wedge\ldots\wedge {\bf dx}^5$ and .
The two operators $\CO_{\CJ F F}$ and $\CO_{\CJ \CJ \CJ}$ account for the ABJ anomaly and the 't Hooft anomaly of the axial current respectively.

 {An important feature of the calculation is that  the Chern-Simons operators appearing in the  fermion determinant are all proportional to the source field $\CJ$; this is because the boundary theory we are describing is free of gauge anomalies. In general, precisely for theories where the boundary theory is free of gauge anomalies, the Chern-Simons contribution to the imaginary part of  $\ln\Delta(B)/\Delta^*(B)$ vanishes when sources vanish, and therefore the fermion path integral only depends on the 4d boundary values of the gauge fields.  This opens the possibility that this 5d construction can behave like a conventional 4d gauge theory.} 

 At finite bulk gap $m$, the $\delta$- and $\theta$-functions in \eq{COs} would be smooth functions  characterizing the normalized profile of the domain wall mode and its integral over  to $r$ respectively, which would be difficult  to compute precisely, but they become  perfectly localized in the infinite  $m$ limit and defined at the  boundary by\footnote{For example, these $\delta$ and $\theta$ can  be regulated as the infinite $m$ limit of the functions $\delta_m = m e^{-m(R-r)}$, $\theta_m =  1- e^{-m(R-r)}$.}
 \beq
 \int_{R-\epsilon}^{R} \delta(R-r) \,dr = 1\ ,\qquad \theta(R-r)=\begin{cases} 1 & r<R\\ 0 & r=R \end{cases}\ ,\qquad \delta(R-r) = -\partial_r \theta(R-r)\ .
\eqn{dtdef}
\eeq

 To obtain \eq{Left} we have integrated by parts and discarded the surface terms, justified by the vanishing of $\theta(0)$.  
  Differentiating $\CL$ with respect to the source $\CJ$ yields,  the conserved current in the effective theory,
 
\beq
J_\mu=  {\frac{\partial\CL_{\text{eft}}}{\partial \CJ_\mu(x)}} &=& \delta(R-r)  \bar q \gamma^\mu \gamma_5  q- \frac{3i\kappa}{2}\theta(R-r) \epsilon_{\mu ijk\ell} \Tr[F_{ij} F_{k\ell}]\ ,\qquad \mu = 1,\ldots 4\cr &&\cr
J_5 =   {\frac{\partial\CL_{\text{eft}}}{\partial \CJ_5(x)}} &=&-\frac{3i\kappa}{2}\theta(R-r) \epsilon_{5 ijk\ell} \Tr F_{ij}F_{k\ell}\ ,
\eqn{cur5d}
\eeq
where $F_{ij} = F_{ij}^a T_a$ and hermitian $T_a$.
Conservation of the 5d current $\partial_k J_k = 0 $ yields
\beq
\partial_\alpha \,\bar q \gamma^\alpha \gamma_5  q = \frac{1}{16\pi^2} \epsilon_{5 \mu\nu\rho\sigma} \Tr F_{\mu\nu}F_{\rho\sigma}\Biggl\vert_{r=R}\ 
\eqn{WTI4d}\eeq
making use of the value of $\kappa$ in \eq{Left}.  In this way the Ward-Takahashi (WT) identity of the exactly conserved $U(1)$ ``isospin''  current in the bulk theory with two Dirac fermions becomes the conventional anomalous  WT identity for the $U(1)_A$ symmetry of the boundary theory with a single massless Dirac fermion, the Callan-Harvey result.

 {We are now in a position to reproduce the GS argument.  We have found the fermion determinant $\Delta(B)/\Delta^*(B)$ to depend only on the boundary gauge field $A$ in the absence of source $\CJ$, and that with the source  the phase is given perturbatively by the Chern-Simons operators  which correctly reproduce the 4d mixed gauge-axial symmetry anomaly and the $U(1)^3$  't Hooft anomalies.  Subsequent integration over   the boundary $SU(3)$ gauge fields should therefore give conventional $N_f=1$ QCD in 4d. }  In particular,   following 't Hooft's argument for a solution to the $U(1)_A$ problem \cite{t1976symmetry} we can model the symmetry structure of the low energy theory by summing over fermion zeromode contributions in the dilute instanton gas model and then matching to  {an effective} Lagrangian\footnote{Chiral symmetry breaking in finite volume requires introducing a small quark mass for the boundary fermions and taking the appropriate limit that this mass goes to zero as the volume goes to infinity.  We discuss this further in the present context in the next section. An effective Lagrangian for the $\eta'$ is motivational at best, unless one invokes a $1/N_c$ expansion,  precisely because it is not an approximate Nambu-Goldstone boson and there is no other parametrically small parameter to expand in.  Nevertheless, considered as a toy model for the $\eta'$, our Lagrangian serves to explain its qualitative features.}.  The  conventional calculation yields
\beq
e^{-S_\text{inst.}}=\sum_{n,\bar n=1}^\infty \frac{\left(\Lambda\int  \bar  q_R  q_L(x)\, dx\right)^n}{n!}  
\frac{\left(\Lambda \int \bar  q_L  q_R(y) \,dy\right)^{\bar n}}{\bar n!}  = e^{\Lambda\int  \left( \bar  q_R  q_L(x) + \bar q_L q_R(x)\right)\,dx}\ ,
\eqn{Isum}
\eeq
where $\Lambda$ is some nonperturbative QCD mass scale inserted by hand to give the correct dimension.  We can now match onto the effective theory of the $\eta'$ meson with the replacement
\beq
 \bar  q_R  q_L(x) \to \frac{\Sigma}{2} e^{i\eta'(x)/f}\ ,\qquad 
 \bar  q_L  q_R(x)\to \frac{\Sigma}{2}e^{-i\eta'(x)/f}\ ,\qquad  
 \frac{\Lambda \Sigma}{ f^2} = M_{\eta'}^2\ ,
\eqn{qsub}
\eeq
 where $\Sigma=\vev{\bar q q}$ is the quark condensate. The resultant effective theory for the $\eta'$ is then
 \beq
 \CL_{\eta'} = \half  \partial_\mu\eta' \partial_\mu\eta'   - M_{\eta'}^2 f^2 \cos\frac{\eta'}{f}\ ,
\eqn{leta} \eeq
 describing an $\eta'$ boson with mass $M_{\eta'}$.   
 
 Obviously something is wrong with the above analysis because the exact  {$U(1)_A$} symmetry in the 5d theory is not realized in \eq{leta}, neither linearly nor nonlinearly as a shift symmetry of the $\eta'$.
Yet this effective theory that follows, via 't Hooft's analysis of the $U(1)_A$ problem plus the assumption that the boundary quark zeromodes are the only low energy degrees of freedom, from   the anomalous Ward-Takahashi identity we derived in \eq{WTI4d}.  Since 't Hooft's analysis is accepted to be qualitatively correct, and  \eq{WTI4d} is  completely conventional for the QCD boundary theory, there would seem to be something wrong with the assumption that the boundary zeromodes are the only light degrees of freedom.

The above conclusion is bolstered by the observation that already in \eq{Isum},  the 't Hooft operators we sum  violate the $U(1)_A$ symmetry of the boundary theory, which is generated by the exactly conserved $U(1)$ bulk current we coupled to the source $\CJ$ in \eq{eft}. These operators were justified as arising from the quark zeromodes in the presence of instantons in the boundary gauge field -- zeromodes which are mandated by the anomalous divergence found in \eq{WTI4d}.  Yet for any 4d world that is the boundary of a 5d manifold, the Dirac index must vanish  when  the entire boundary is included, so these zermodes cannot be the whole story. Given that a single instanton configuration can be placed on the outer boundary of the disk, and that an observer on that boundary should see a chiral fermion zeromode in its vicinity, it follows that the manifold must possess another boundary with a zeromode of the opposite chirality in order to obey the index theorem.  Thus a new boundary must have appeared--which can take the form of  a singularity in the gauge field in the bulk--with mirror zeromodes located there\footnote{On the lattice there would be dislocations in the gauge field rather than a singularity, and the bulk zeromode could arise on the boundary of a hypercube surrounding the dislocation. }. 

 {The error in our chain of reasoning therefore lies with the assumption that the bulk fermions will be fully gapped.  While this will be true in the presence of gauge fields that are smooth on the scale of the bulk fermion mass, singularities in the gauge field can -- and, in this example, must -- result in fermion zeromodes in the bulk. Such zeromodes will be highly localized, falling off from the instanton exponentially fast rather than with the familiar power law.}

  With  {localized} mirror zeromodes appearing in the bulk, the conventional anomaly inflow argument is incomplete since the effective Lagrangian in \eq{eft} and the Ward identity derived from it failed  to include the mirror fermion contributions, as did the 't Hooft operators we summed  in \eq{Isum}.  When the calculation is done correctly in the presence of boundary instantons,  bulk zeromodes will cause bulk fermions to appear in the 't Hooft operators as well as the boundary quark fields, as discussed in \cite{grabowska2016nonperturbative,golterman2024conserved}, allowing the  exact $U(1)$ symmetry to be respected.  At this point we are in agreement with the GS analysis.  However, before we consider how to modify \eq{Isum}, we first must consider the structure of the bulk gauge field in more detail  {in order to understand how gauge fields with nontrivial topology might be continued into the bulk.}

%%%%%%%%%%%%%%%%%
 
\section{Engineering the bulk gauge field}

%%%%%%%%%%%%%%%%%
 {We have seen that when the boundary theory does not suffer from gauge anomalies, the (naive) computation of the fermion determinant $\Delta(B)/\Delta^*(B)$ yields a result that only depends on the gauge fields at the boundary.  It is tempting then to conclude that it doesn't matter how the 5d gauge fields are continued into the bulk, and they could be set to zero, modulo 4d gauge transformations.  The problem with this is that the existence of light edge states depends on the nontrivial topological phases that result when the bulk fermions are fully  gapped.  They are gapped in the free theory, and thus will also be when coupled to fields that are smooth at the scale of the gap, but there is no reason for this to hold when such a smoothness condition is violated.  We will assume that the 4d theory is asymptotically free and that the 4d fields at the boundary are sufficiently smooth at the UV cutoff scale set by the the gap, while the continuation of the fields away from the boundary into the extra dimension must also be smooth\footnote{Smoothness in a gauge theory must be defined for gauge invariant quantities; on a lattice one will always have to see whether rough gauge fields spoil the desired physics, although it is reasonable to hope that when close enough to the continuum limit and with an improved action, such field configurations are rendered unimportant.}.  }

 {In order to ensure a smooth continuation of gauge fields into the bulk in the vicinity of the boundary, it was suggested in Ref.~\cite{kaplan2024chiral} that   the 5d gauge field $B_k$ be defined as the solution to the Yang-Mills equations subject to the boundary condition 
\beq B_\mu(x)\bigl\vert_{\delta X} = A_\mu(\bfx)\ ,\qquad B_5(x)\bigl\vert_{\delta X} =0.
\eeq 
More conventional gradient flow might also be a possible continuation, although it seems that could lead to unnecessary singularities at the center of the disk.

 {Having the bulk gauge fields be smooth near the boundary protects the light edge states from violent gauge field fluctuations, but it is in general impossible to design a 5d completion of the boundary gauge field that is smooth everywhere in the bulk.  For example, if the 4d boundary field has nontrivial winding number and is continued smoothly into the bulk, the gauge field on a 4d surface lying just within the boundary would have to possess the same winding number.  As one retreats into the bulk by contracting this surface, that topology will have to change abruptly at some point before the surface can be contracted to a point.  Such a change requires a singularity in the gauge field somewhere in the bulk -- albeit, far from the boundary if one defines the flow of the gauge fields from a smooth boundary configuration via the equations of motion.}

 { Once one admits bulk gauge configurations that can have singularities, it is no longer evident that the theory remains gapped in the bulk, calling into question the Callan-Harvey analysis of the previous section where one integrates out the bulk fermions.  Furthermore, there is nothing universal about the number and locations of the singularities, which will depend not only on the boundary gauge field, but also on the exact implementation of the gauge field flow used to define the bulk gauge field.}

For simplicity we will use the dilute instanton gas model to  {illustrate the effects of} nontrivial gauge topology. {Furthermore, since gauge field singularities effectively create disconnected boundaries in the bulk, it is convenient to start by discussing an annulus rather than a disk (this discussion is similar to that for a 5d strip with two boundaries discussing in Ref.~\cite{grabowska2016nonperturbative})}, the inner boundary representing more generally the boundary created by gauge field singularities deep withing the bulk.  If the  {outer} boundary configuration includes $n$ instantons and $\bar n$ anti-instantons, then since this configuration is not an exact solution to the 4d Euclidian equations of motion, we would expect instantons and anti-instantons to flow together and annihilate pairwise as the gauge field evolves into the bulk.  If the flow continues long enough before reaching the inner boundary of the annulus and encounters no gauge field singularities, it is plausible that one is left with only the minimal instanton configuration on the inner boundary required for winding number $\nu=(n-\bar n)$.  {We refer to this type of flow as {\it annealing flow}, and assume that it can be realized on the lattice:  maximally efficient damping out of local topological features in the bulk, consistent with global topological constraints.}  
 {Such flow will give rise to $\nu$ fermion zeromodes on the inner boundary;
we will refer to these zeromodes as $Q$ fields, to distinguish them from $q$ on the outer boundary. They must be included in the low energy effective theory after the gapped bulk modes have been integrated out. Our assumption of annealing flow differs from from the equally valid flow GS assumed, which is  responsible in part for why we reach different conclusions.  Whether annealing flow can be achieved in a realistic lattice simulation is something that needs to be explored; studies of instanton cooling suggest that it may be possible \cite{perez1994instantons,de1997topology}.

%%%%%%%%%%%%%%%%%

\section{Resolution of the $\mathbf{U_A(1)}$ problem}

%%%%%%%%%%%%%%%%%

{We can now redo the analysis in \S\ref{U1A}, assuming the geometry of an annulus and  taking into account the effect of the $Q$ zeromodes on the inner boundary,  making use of 't Hooft's qualitative analysis of the $U(1)_A$ problem in the context of the dilute instanton gas model.}   {We will work at finite but large volume $V$ and begin with the theory of \eq{eft} with its exact $U(1)\times U(1)$ symmetry.  
The  assumption of annealing flow for defining the bulk gauge fields ensures that
(i) the only gauge configurations that survive to the deep interior are the minimal number $|\nu|$ instantons or anti-instantons required by topology preservation; and (ii)
the spatial locations of the surviving $|\nu|$ instantons or anti-instantons
at the inner boundary are decorrelated with the spatial locations of the $|n+\bar n|$ instantons and anti-instantons on the outer boundary, due to the nonlinearity of the flow and typically large $n$ and $\bar n$ appearing in large volume at the outer boundary.}

 {The analog of \eq{Isum} now involves both $q$ and $Q$ fields,
\beq
e^{-\tilde{S}_{\text{int}}}\equiv Z_\text{topo}&=&\sum_{n,\bar n}  \frac{(V \CO)^n}{n!} \frac{(V\bar \CO)^{\bar n}}{\bar n!} \left( (V'  X)^{(n-\bar n)} \Theta(n-\bar n)  +  (V'  \bar X)^{(\bar n-n)} \Theta(\bar n-n) +\delta_{n,\bar n}\right)\cr &&\cr 
&=& Z_1+Z_2+Z_3
\eqn{isum}
\eeq
with $V$ and $V'$ being the 4-volumes of the outer and inner boundaries respectively, $\Theta$ defined as 
\beq
\Theta(n) = \begin{cases} 1 & n> 0\\ 0 & n\le 0 \end{cases}\ ,
\eeq
and the dimension 4 operators given by
\beq
\CO =  \Lambda\int_V \frac{d^4x}{V}\,  \bar q_{R} q_{L} \ ,\quad
\bar \CO=  \Lambda\int_V \frac{d^4x}{V}\,  \bar q_{L} q_{R} \ ,\quad
X = \Lambda\int_{V'} \frac{d^4y}{V'}\,  \bar Q_{R} Q_{L} \ ,\quad
\bar X =  \Lambda\int_{V'} \frac{d^4y}{V'}\,  \bar Q_{L} Q_{R} \ ,
\eqn{ops}
\eeq
where $\Lambda$ is an infrared scale put in by hand, by dimensional analysis.  For simplicity we assume without justification the same scale for both $\CO$ and $X$ operators, but this assumption will not affect our analysis.  For $\CO$ and $\bar \CO$ the $x$ integral is over the outer surface with 4-volume $V$, while for $X$ and $\bar X$ the $y$ integral is over the inner surface with 4-volume $V'$, which is assumed to be fixed and small as $V$ gets large. }

 {The action of the $U(1)\times U(1)$ symmetry  on the 5d Dirac fermions $\psi^\pm$ and their associated edge states $q$ and $Q$ is given by
\beq
\psi^+\to e^{i\alpha}\psi^+:\ \ \{q_L\to  e^{i\alpha} q_L\ ,\ \ Q_R\to e^{i\alpha}Q_R\}\ ,\qquad
\psi^-\to e^{i\beta}\psi^-:\ \  \{q_R\to  e^{i\beta} q_R\ ,\ \ Q_L\to e^{i\beta}Q_L\}\ ,
\eeq
so that the operators $\CO$ and $X$ are invariant when $\alpha=\beta$ (interpreted as a $U(1)_V$ transformation in the boundary theory), while under the $U(1)_A$ transformation with $\alpha = -\beta$ they transform as
\beq
\CO\to e^{2i\alpha}O\ ,\quad \bar\CO\to e^{-2i\alpha}\bar O\ ,\quad X\to e^{-2i\alpha} X\ ,\quad \bar X\to e^{2i\alpha} \bar X\ .
\eeq
The instanton sum in \eq{isum} is therefore properly $U(1)\times U(1)$ invariant, like the underlying theory and unlike the naive attempt in \eq{Isum}.}

If we are only interested in the boundary physics -- e.g. correlation functions that involve only the boundary fermions $q$ -- then  we can restrict our attention in \eq{isum} to just the $Z_3$ term, where
\beq
Z_3\equiv e^{-\widetilde S_\text{inst.}}&=&
\sum_{n=0}^\infty \frac{\left(\Lambda\int  \bar  q_R  q_L(x)\, dx\right)^n}{n!}  
\frac{\left(\Lambda \int \bar  q_L  q_R(y) \,dy\right)^{ n}}{  n!} 
= I_0\left( 2\Lambda V \sqrt{\overline{\bar  q_R  q_L}\ \overline{\bar  q_L  q_R}}\right)
\ .
\eqn{Isum2}
\eeq
In this expression $I_0$ is a modified Bessel function and the bar notation indicates an average over Euclidian  spacetime,
\beq
\overline{\CO} \equiv \int \frac{d^4x}{V} \, \CO(x)\ .
\eeq
In the large-$V$ limit the expression for $\widetilde S_\text{inst.}$ in \eq{Isum2} simplifies to
\beq
 \widetilde S_\text{inst.}\xrightarrow[]{V\to\infty}
 V\left[- 2\Lambda  \sqrt{\overline{\bar  q_R  q_L}\ \overline{\bar  q_L  q_R}} +O\left(\frac{\ln V}{V}\right)\right]\ .
\eeq
{Below the chiral symmetry breaking scale we can use the substitutions in \eq{qsub} to model the theory of the $\eta'$ as }
\beq
\widetilde S_{\eta'} &=& -V M_{\eta'}^2 f^2  \sqrt{\overline{\left(e^{i\eta'/f}\right)}\ \overline{\left(e^{-i\eta'/f}\right)}}+\int  \half  \partial_\mu\eta' \partial_\mu\eta'(x) \, d^4x\ ,
\eeq
where as before $M_{\eta'}^2 f^2=2\Lambda \Sigma$. This can be represented by a nonlocal Lagrange density
\beq
\widetilde \CL_{\eta'} &=&
 \half  \partial_\mu\eta' \partial_\mu\eta'(x)   + M_{\eta'}^2 f^2\left[ -1 + \frac{\left(\eta'(x) - \overline{\eta'} \right)^2}{2f^2} + O(\eta'^4 )\right]
\ .
\eqn{Leta}
\eeq
  This  nonlocal action $\widetilde S_{\eta'}$ evidently possesses two desirable features which are not mutually compatible   in local theories: (i) the $\eta'$ is massive, and (ii) the theory possesses the exact $U(1)_A$ symmetry under which the $\eta'$ field transforms as $\eta'(x)/f \to \eta'(x)/f + \alpha $, where $\alpha$ is an arbitrary $x$-independent angle. This is possible because of the existence of an isolated zeromode at $p=0$, the $\overline{\eta'}$ degree of freedom\footnote{ This action $\widetilde S_{\eta'}$ is identical to the effective action for the irrational axion  in Ref.~\cite{banks1991irrational} after integration over $\theta$.}.

 {The remaining contributions in \eq{isum} involving bulk zeromodes via the $X$ operators will not contribute to vacuum matrix elements of operators only involving fields on the outer boundary. Because of the gap in the bulk, the $Q$ zeromodes are highly localized and do not propagate to the outer boundary.   
 Furthermore, the $X$ operators will not develop vacuum expectation values because there cannot be spontaneous symmetry breaking on the inner boundary with its small  volume, which is kept finite as $V\to\infty$. }

We have shown how chiral symmetry breaking and a massive $\eta'$ meson in the boundary spectrum can be consistent with the existence of an exact $U(1)_A$ symmetry, at least with the assumption of annealing flow for the bulk continuation of the boundary gauge fields. {However, we have not addressed the specific spectral argument of  Ref.~\cite{golterman2024conserved}   where GS constructed a spontaneously broken and exactly conserved 4d $U(1)$ current by integrating a 5d current over the fifth dimension, and argued -- in the spirit of  proofs of the Goldstone theorem -- that spontaneous breaking of the symmetry generated by the current was inconsistent with a gapless spectrum.}  In particular they considered the WT identity
\beq
\partial^x_\mu \vev{J_\mu(x)P(y)} = \delta(x-y) \Sigma\ ,
\eqn{GSWT}
\eeq
where $J_\mu$ is their 4d exactly conserved and gauge invariant $U(1)$ current,  $P(y) = \bar q i \gamma_5 q(y)$ is the pseudoscalar quark density, and  $\Sigma$ is the vacuum expectation value of the scalar quark density.  The Fourier transform of this equation evidently cannot hold as $p\to 0$ unless 
$\vev{J_\mu(p)P(-p)}$ is proportional to $p_\mu/p^2$, which they argued mandates the existence of a Nambu-Goldstone boson in the spectrum. If our above reasoning is correct for \bcf, then there must be a flaw in their argument.

It is instructive to construct the analogous WT identity for the exact $U(1)_A$ symmetry of the effective $\widetilde \CL_{\eta'}$ theory of \eq{Leta} to understand how it satisfies Goldstone's theorem.  By performing the change of variables $\eta'(x)/f \to \eta'(x)/f + \alpha(x) $ and setting to zero the functional derivative of the partition function with respect to $\alpha(x)$ we generate the WT identity for this theory
\beq
- \Sigma \,\partial_\mu\vev{\partial_\mu\eta'(x)\eta'(y)}
+ M_{\eta'}^2 \Sigma\vev{\left(\eta'(x)-\overline{\eta'}\right) \eta'(y)}
%+\frac{m_q\Sigma^2}{f^3}\langle\eta(x)\eta(y)\rangle
 =  \Sigma  \delta^4(x-y)\ .
\label{wti3}
\eeq

The above identity looks problematic if one integrates over $x$ and $y$, the left side integrating to zero while the right side integrates to $V\Sigma$,
which seems to validate the Golterman-Shamir argument.  However, in this case the problem is due to the familiar fact that one cannot see spontaneous symmetry breaking in finite volume.  To analyze the model correctly  {we must reintroduce a small explicit symmetry breaking term -- such as a quark mass $M_q$  {localized at the outer boundary}} -- and then take the combined chiral and thermodynamic limits with $M_q L \to 0$, $M_q \Sigma V\to \infty$, where $V = L^4$ \footnote  {This discussion follows reasoning similar to that found in Ref.~\cite{leutwyler1992spectrum}.}. {The addition of small real quark mass modifies our effective Lagrangian in \eq{Leta}, which now becomes at leading order in $M_q$}
\beq
\widetilde \CL_{\eta'}(M_q) &=&
 \half  \partial_\mu\eta'(x) \partial_\mu\eta'(x) - M_q \Sigma\cos\frac{\eta'(x)}{f}  - M_{\eta'}^2 f^2\left[ 1 -\frac{ \left(\eta'(x) - \overline{\eta'} \right)^2}{2f^2} +\ldots \right]
\cr && \cr
 &=&
 \half  \partial_\mu\eta'(x) \partial_\mu\eta'(x) +\frac{M_q \Sigma}{2f^2}\left(-2f^2 +\eta'(x)^2+O(\eta'^4/f^2)\right)  - M_{\eta'}^2 f^2\left[ 1 -\frac{ \left(\eta'(x) - \overline{\eta'} \right)^2}{2f^2} +\ldots \right]
\ ,
\eqn{Leta2}
\eeq
giving rise to the modified WT identity
\beq
- \Sigma \,\partial_\mu\vev{\partial_\mu\eta'(x)\eta'(y)}
+ M_{\eta'}^2 \Sigma\vev{\left(\eta'(x)-\overline{\eta'}\right) \eta'(y)}
+\frac{M_q\Sigma^2}{f}\vev{\eta'(x)\eta'(y)}
 =  \Sigma  \delta^4(x-y).
\label{wti4}
\eeq
Now on integrating over $x$ and $y$ one obtains
\beq
V^2\frac{M_q\Sigma^2}{f^2}\vev{\left(\,\overline{\eta'}\,\right)^2} =  \Sigma V\ .
\eeq
The $\overline{\eta'}$ degree of freedom appears in $\widetilde \CL_{\eta'}(M_q)$ only in the $M_q$ term and is seen to be a Gaussian variable with $\vev{\left(\overline{\eta'}\right)^2} = f^2/(M_q\Sigma V)$, and so we see that it saturates the WT identity properly, in a way that persists as $M_q\to 0$ and explicit symmetry breaking is removed (simultaneously taking $V\to\infty$ such that $M_q\Sigma V$ is large enough to justify expanding $\CL$ to quadratic order in $\overline{\eta'}$).

We see that this model meets the minimum requirement of a theory which has a spontaneously broken exact symmetry:  there must be a bosonic field with an exact shift symmetry.  However, since the theory is not strictly local, it does not possess a conventional massless Nambu-Goldstone boson as claimed in ref.~\cite{golterman2024conserved}.  Rather, the $p_\mu=0$ mode of the $\eta'$ possesses the exact shift symmetry, while for $p_\mu\ne 0$, the fluctuations of the $\eta'$ look like that of a conventional massive boson.  Since any realistic experiment will always take place in finite volume, there would be no experimental evidence in this world for a massless meson.

% The lesson this model teaches us is  that  when one has a spontaneously broken global symmetry, the requirement is not that there must be a \textcolor{red}{Goldstone} boson, but simply that the effective theory must  realize the symmetry nonlinearly as a shift symmetry. \textcolor{red}{This allows for the possibility that the theory can be gapless or massless, but not necessarily linearly dispersing}. The shift symmetry requires that there must be an isolated mode with $p_\mu = 0$; in a local theory this is synonymous with having a propagator proportional to $1/p^2$, i.e. linearly dispersing massless boson, and the corresponding long-range correlations in spacetime. We see that in the nonlocal theory we are studying, however, the $p_\mu = 0$ mode of the $\eta'$ is isolated, and there is no \textcolor{red}{linearly dispersing} mode in the spectrum. As a result there are no power law correlations in the shifted field $(\eta' -\langle \eta' \rangle)$. \textcolor{red}{Thus, while we agree with GS that the low energy theory is massless in the sense of there being an isolated $p=0$ mode for the $\eta'$ that is ungapped, we show that our construction does not result in a linear dispersion for $\eta'$ as claimed by GS.} 

 {Our argument has not addressed the specific current devised by GS in \eq{GSWT},   a 4d current derived by integrating with respect to $r$ the first four components of a 5d bulk current, which they showed was exactly conserved.  We argue that this is not a reliable current to use when concluding that a $1/p^2$ pole must be generated by a massless particle to saturate the identity.  This is because it is a nonlocal object which invalidates the conclusions.  The GS current can be constructed directly from the conserved 5d current in \eq{cur5d} following their procedure of integrating over the extra dimension, and that current depends on the bulk gauge fields from  the Chern-Simons operator.  In the \bcf ~construction those bulk gauge fields depend nonlocally on the dynamical gluon field at the boundary via the gapless Green function for the classical Yang-Mills field.   Thus we should not be surprised to find poles at $p=0$ in its matrix elements as they are built into the operator;  their existence does not require a massless boson in the spectrum.  
  In fact it is easy to construct an example of such a current in the effective theory of the $\eta'$.  The equation of motion for the $\eta'$ in that theory is of the form
\beq
\partial^2 \eta'(x)+ \CA(x)=0\ . 
\eeq
The exact expression for $\CA(x)$ is easily computed in terms of the $\eta'$ field but its structure is not particularly interesting.)  From this equation one can construct a nonlocal, exactly conserved current which acts nonlinearly on the vacuum (and which is gauge invariant, of course, being constructed of color neutral mesons): 
\beq
j_\mu = \partial_\mu \eta'  +  \frac{\partial_\mu}{\partial^2}\CA\  \ .
\eeq
This is a candidate for the nonlocal GS current in the effective theory, conserved and gauge invariant, but which   seems to have  little physical significance and which coexists happily in a theory without a massless boson. The GS current could plausibly have this structure in the effective theory, given how the bulk gauge fields it contains are nonlocally defined.}

%%%%%%%%%%%%%%%%%

\section{A new solution to the strong $CP$ problem?}

%%%%%%%%%%%%%%%%%

 {We have argued that while the \bcf ~model differs from conventional formulations of 4d QCD in that it possesses an exact $U(1)_A$ symmetry, the phenomenology in the $\eta'$ sector can be indistinguishable from that of QCD.  In particular, so long as we -- inhabitants of the 4d world on the outer boundary -- cannot access matrix elements of operators involving fields in the bulk, then the phenomenology resembles a world where the topological charge has been constrained to equal zero.  There has been no such constraint on the theory, but the bulk zeromodes dynamically kill all contributions to the path integral at nonzero winding number, and to all Green functions accessible to an experimentalist confined to the outer 4d boundary of the 5d world.}

 {A consequence of the bulk zeromodes is that there cannot be any strong $CP$ violation in the boundary theory.  A direct way to see this is to include in the action a complex quark mass on the outer boundary,
\beq
\int d^5x\, \delta(r-R)\left[M_q e^{2i\theta} \bar\psi^+\psi^- + h.c.\right]\to \int d^4x\, \left[M_q e^{2i\theta} \bar q_Lq_R + h.c.\right]\ .
\eeq
By rotating $\psi^\pm \to e^{i\theta}\psi^\pm$ one can remove the angle $\theta$ entirely from the theory, provided that there is no analogous quark mass in the region where the bulk zero modes appear, far from the outer boundary. There is no analog of the chiral anomaly in 5d which would transfer the $CP$ violating phase to the gauge sector. Thus  the phase  of the quark mass $M_q$ is unphysical (or, with more than one flavor, the phase of the determinant of the quark mass matrix is unphysical), and there cannot be any strong $CP$ violation. The model resembles  a version of conventional  QCD with  a massless  up quark, allowing the phase of the quark mass determinant to be rotated away.  However, in \bcf, the role of the massless up quark and its zeromodes in the presence of instantons is played by the bulk zeromodes -- except that the bulk fermions do not affect hadron phenomenology, unlike a massless up quark, which is found  in lattice QCD computations to be inconsistent with the  observed hadron spectrum \cite{aoki2025flagreview2024}.}

 It is important to emphasize that our analysis relies on the quark mass term $M_q$ being inserted only on the boundary of the five-dimensional world.  If we had $U(1)_A$-violating $M_\psi$ mass term that was constant over the entire fifth dimension, the resulting boundary QCD theory would be expected to have a conventional strong $CP$ problem. That the two constructions differ would seem to be a violation of the concept of universality, and it arises because the theory in our construction admits zeromodes in the bulk which we cannot experimentally access with boundary degrees of freedom due to the gap in the bulk.
 
 {It is interesting that while there is no obstacle in principle to constructing a conventional 4d version  of lattice QCD with strong $CP$ violation, when QCD is embedded in the Standard Model chiral gauge theory, the \bcf ~proposal with only boundary quark masses produces a regulated version of QCD without strong $CP$ violation -- which is consistent with what we actually see in the world. It is natural to wonder then whether this mechanism could provide a viable solution to the strong $CP$ problem.  An unusual feature of \bcf ~is that it is a 5d theory without dynamical gauge fields in 5d; whether that can be realized in a theory in Minkowski spacetime remains a question to explore.}

\section{Discussion}

{Although \bcf ~was constructed for regulating chiral gauge theories, in this paper we have followed the lead of Ref.~\cite{golterman2024conserved} and have focused on an example where the boundary theory looks like QCD, a vector-like $SU(3)$ gauge theory with massive Dirac quarks.  Of course, much more practical methods exist for simulating QCD on the lattice in the absence of  weak interactions!  Those methods can in principle simulate $SU(3)$ gauge theory with any moderate number of quark flavors, with any possible $CP$-violating $\bar\theta$ parameter, even if there are practical obstacles to a simulation at nonzero $\bar\theta$.  It is conventional to consider such a theory to be identical to the strong interaction sector of the Standard Model -- namely that the simulation of the stand-alone vector-like $SU(3)$ theory will be just ``plugged into'' the simulation of the full chiral gauge theory, once the pesky details of such a simulation are ironed out.  However, without a concrete, working lattice regulator for chiral gauge theories, this must be recognized as an assumption  that might not necessarily be  valid.  It is possible that QCD embedded in the full Standard Model with its chiral charges does not behave the same as a QCD-like theory by itself.  In the context of a strictly four-dimensional lattice construction, this possibility seems like a far-fetched violation of universality. In this paper we have suggested that a concrete realization of such a failure of 4d universality  might occur when simulating a chiral gauge theory requires an extra dimension.  Of course, stand-alone simulations of $SU(3)$ gauge theory match the observed hadronic spectrum and properties to high accuracy, so any distinction between such a theory and QCD as it appears in the Standard Model must be subtle.  Here we suggest that observable  distinctions would only involve topological effects and the anomalous $U(1)_A$ symmetry  such that the QCD we observe in the real world does not look like a generic $SU(3)$ gauge theory with quarks, but can naturally conserve $CP$ symmetry  without constraining the phases in the quark mass matrix or having a light $0^{-+}$ boson in the spectrum.}

 {The specific goal of this paper was to address critiques of the proposal from Refs.~\cite{grabowska2016nonperturbative,kaplan2024chiral,kaplan2024weyl}   arising in Ref.~\cite{golterman2024conserved} -- which deals with the existence of unexpected exact global $U(1)$ symmetries of the 5d theory  -- and in Refs.~\cite{aoki2022curved, aoki2023curved,aoki2023magnetic, aoki2024,aoki2024study}, -- which point out the prevalence of fermion zeromodes in the bulk of the fifth dimension in the presence of certain boundary gauge fields. Both of these phenomena would seem to interfere with a boundary theory emerging with conventional phenomenology.  How these phenomena manifest themselves in the boundary theory depend on how the gauge field is continued into the bulk, however.  In this paper we assumed that (i) the gauge field continued smoothly into the bulk near the boundary (near in units of the bulk fermion mass), and (ii) that far from the boundary, topological defects were ``annealed'', with only the minimally required defects surviving deep into the bulk.  We advocated using the Euclidian equations of motion to define the flow, since they will favor the annihilation of instanton/anti-instanton pairs.  On the lattice, the efficiency of such flow to anneal the gauge field will depend on the specific discretization of the flow equations \cite{perez1994instantons,de1997topology}, and it remains to explore how this might work on the proposed lattice geometry.  With these assumptions we argued that when the boundary gauge topology was nontrivial, fermion zeromodes would appear deep in the bulk, exponentially localized.  Thus the existence of such zeromodes will eliminate the contribution of boundary gauge fields 
with nontrivial topology to matrix elements of operators living solely on the boundary.  Whether this scenario can be realized in practice is something which could be explored in computations without any obvious insurmountable obstacles, such as a bad sign problem.}

 {While the bulk zeromodes far off in the fifth dimension act as cosmic censors which ensure that we can only access trivial gauge field topology experimentally, they do not suppress local topological fluctuations.  Thus the dispersion relation for the $\eta'$ at nonzero momenta is that for a massive particle; its $p=0$ mode, however, is an exact zeromode of the theory. One similarly finds conventional results for the closely related correlator of topological fluctuations in the gluon field, $K(p) = \int d^4x\,e^{ipx} \vev{\tilde GG(x) \,\tilde G G(0)}$, for all nonzero values of $p_\mu$.   As a result,  the theory can possess an exact global $U(1)_A$ symmetry that is spontaneously broken without a corresponding Nambu-Goldstone mode -- something that is impossible in a purely four-dimensional theory.  That $U(1)_A$ symmetry, which is exact in the full theory but anomalous on the boundary, allows one to rotate into bulk degrees of freedom any $CP$-violating phase of the boundary theory, rendering it unphysical in the boundary theory.  \bcf~therefore provides an example for how QCD as part of a chiral gauge theory (the Standard Model) might look different from a purely 4d vector-like gauge theory with the same particle content.  Since the difference we are finding addresses an   outstanding problem of the Standard Model -- the strong $CP$ problem -- perhaps this exotic conclusion should be taken seriously.
}

 {While the \bcf~construction was intended as an approach for computing the consequences of chiral gauge theory on a computer, if a consequence is the resolution of the strong $CP$ problem it is natural to ask whether this might be how the universe actually works, e.g., whether we consist of edge states in a 4d world that forms the boundary of a higher dimensional spacetime which behaves like a topological insulator.  For  a dynamical, cosmological model of this sort to be considered, one would have to be able to construct a version of the theory in Minkowski spacetime. A central challenge for such a construction is understanding how one can realize its 5d gauge fields in real time that exhibit only 4d dynamics.}

\vspace{1in}
\section{Acknowledgement}
We thank Claudio Bonanno, Aleksey Cherman, Maarten Golterman, Y. Kikukawa, Carlos Pena, Yigal Shamir, Mithat \"Unsal, U. J. Weise and Larry Yaffe for useful conversations. 
DBK would like to thank the long-term workshop on HHIQCD2024 at the Yukawa Institute for Theoretical Physics (YITP-T-24-02) for fruitful discussions and opportunity to complete this work and the Instituto de F\'isica Te\'orica, Universidad Aut\'onoma de Madrid for hospitality. DBK is supported in part by DOE Grant No. DE-FG02-00ER41132 and received partial support from the Severo Ochoa Program,  grant CEX2020-001007-S, funded by MCIN/AEI/10.13039/501100011033. 
We acknowledge   hospitality of the Simons Center during the workshop ``Symmetric Mass Generation, Topological Phases and Lattice Chiral Gauge Theories'', 2024. SS is supported by the U.S. Department of Energy,
Nuclear Physics Quantum Horizons program through the
Early Career Award DE-SC0021892. 

\appendix

%%%%%%%%%%%%%%%%%

\section{The bulk fermion determinant on the lattice}

As discussed in the text, in the continuum the modulus of the bulk fermion determinant can be canceled by the same Pauli-Villars field that regulates the phase of the determinant and gives rise to its topological significance.   In the lattice version, everything is completely regulated by the discretization and the theory can be in a number of different topological phases, depending on the parameters of the theory.  However, we still wish to cancel the modulus of the bulk fermion determinant to remove, for example, a 5d Yang-Mills kinetic term contribution that will interfere with a 4d interpetation of gauge field dynamics.  Here we address how to do that. 

In the continuum we have 5d fields with fermion operator $(\slashed{D}\pm m)$ assuming positive bulk mass $m>0$, with the boundary condition $(1\mp \gamma_r)\psi(R)^\pm=0$, resulting in light edge modes with chirality $\gamma_5 = \pm 1$ in the 4d boundary theory.  
The lattice version of the theory simply involves replacing the Dirac operator $(\slashed{D}\pm m)$ for each 5d bulk fermion of the continuum theory by the Wilson operator   $D_w^\pm = \slashed{D} \pm (1 + \half D^2)$ with open boundary conditions at $r=R$, where here $\slashed{D}$ and $D^2$ are the conventional discretized version of the covariant Dirac operator and Laplacian respectively (in lattice units with $a=1$) 
\cite{kaplan2024weyl,Golterman:1992ub}.  For each lattice fermion with operator $D_w^\pm$ there will similarly be a light boundary mode with chirality $\pm 1$.

To remove the modulus of the bulk fermion determinant we propose  the following replacement in the lattice formulation:
\beq
\det D^{\pm}\to \frac{\det D_w^{\pm}}{\sqrt{\det\left({D_w^{\pm}}^{\dagger}D_w^{\pm}+\mu^2\delta_{r,R}+\epsilon\right)}}.
\eqn{ferm2}
\eeq
If the theory were fully gapped, the first term in the denominator would suffice by itself.  However, since ${D_w^{\pm}}$ has low-lying eigenvalues at the boundary whose contribution to physics we do not want to cancel, we have included a mass term $\mu^2\delta_{r,R}$ in the denominator which is only nonzero at the boundary.  Finally, as discussed in this paper, we expect $D_w^\pm$ to exhibit zeromodes in the bulk whenever the topology of the boundary gauge fields is nontrivial, and we want the fermion determinant to vanish at these points in field space.  Since ${D_w^{\pm}}^{\dagger}D_w^{\pm}$ will also vanish there, and the $\mu^2$ term will not affect the eigenvalues of bulk zeromodes, we include a small term $\epsilon$ to shift the denominator away from zero.  This parameter $\epsilon$ can  be taken very small and will have a vanishing effect on the desired cancellation of the modulus of the unmodified fermion determinant in the numerator of our expression.

We do not have local Pauli-Villars field representation of the denominator for a  single bulk fermion.  However, for an even number of bulk fermions with the same sign mass and the same representation under the gauge group, the square root goes away and a Pauli-Villars implementation is straight forward.  Such an implementation might be the first step toward   a Hamiltonian formulation of the theory for quantum simulation or for realizing the theory in Minkowski spacetime.

\bibliography{refs.bib}
\end{document}

%% file: preamble.tex
\usepackage[T1]{fontenc}
\usepackage{enumitem}
\usepackage{pifont}
\usepackage{amsmath}

\usepackage[mathscr]{euscript}
\makeatletter
\newcommand{\subalign}[1]{%
  \vcenter{%
    \Let@ \restore@math@cr \default@tag
    \baselineskip\fontdimen10 \scriptfont\tw@
    \advance\baselineskip\fontdimen12 \scriptfont\tw@
    \lineskip\thr@@\fontdimen8 \scriptfont\thr@@
    \lineskiplimit\lineskip
    \ialign{\hfil$\m@th\scriptstyle##$&$\m@th\scriptstyle{}##$\hfil\crcr
      #1\crcr
    }%
  }%
}
\makeatother
 \usepackage{mathtools}
 \usepackage{lmodern}
 \usepackage{lmodern}
\usepackage{graphicx}
\usepackage{hyperref}
\usepackage{amsfonts}
\usepackage{tocvsec2}
\usepackage{slashed}
\usepackage{cancel}
\usepackage{comment}

\newcommand\beq{\begin{eqnarray}}
\newcommand\eeq{\end{eqnarray}}

\newcommand{\half}{\frac{1}{2}}

\newcommand\eqn[1]{\label{eq:#1}} 
\newcommand\eq[1]{eq.~(\ref{eq:#1})}

\newcommand{\bfx}{{\mathbf x}}

\newcommand{\CA}{{\cal A}}

\newcommand{\CJ}{{\cal J}}

\newcommand{\CO}{{\cal O}}

\newcommand{\CL}{{\cal L}}

\newcommand{\Tr}{{\rm Tr\,}}

\newcommand\vev[1]{\left\langle #1 \right\rangle}

\newcommand{\mybar}[1]{\kern 0.6pt\overline{\kern -0.6pt#1\kern -0.6pt}\kern 0.6pt}

\def\Tr{\text{Tr}\,}

\def\half{\tfrac{1}{2}}

\def\U\Omega{U(1)_{\Omega}}

\hypersetup{
  colorlinks=true,   % false: boxed links; true: colored links
  linkcolor=cyan,    % color of internal links (box color = linkbordercolor)
  citecolor=blue,    % color of links to bibliography
  filecolor=magenta, % color of file links
  urlcolor=blue      % color of external links
}
\usepackage[section]{placeins}

%% file: submit5.bbl
\begin{thebibliography}{26}%
\makeatletter
\providecommand \@ifxundefined [1]{%
 \@ifx{#1\undefined}
}%
\providecommand \@ifnum [1]{%
 \ifnum #1\expandafter \@firstoftwo
 \else \expandafter \@secondoftwo
 \fi
}%
\providecommand \@ifx [1]{%
 \ifx #1\expandafter \@firstoftwo
 \else \expandafter \@secondoftwo
 \fi
}%
\providecommand \natexlab [1]{#1}%
\providecommand \enquote  [1]{``#1''}%
\providecommand \bibnamefont  [1]{#1}%
\providecommand \bibfnamefont [1]{#1}%
\providecommand \citenamefont [1]{#1}%
\providecommand \href@noop [0]{\@secondoftwo}%
\providecommand \href [0]{\begingroup \@sanitize@url \@href}%
\providecommand \@href[1]{\@@startlink{#1}\@@href}%
\providecommand \@@href[1]{\endgroup#1\@@endlink}%
\providecommand \@sanitize@url [0]{\catcode `\\12\catcode `\$12\catcode
  `\&12\catcode `\#12\catcode `\^12\catcode `\_12\catcode `\%12\relax}%
\providecommand \@@startlink[1]{}%
\providecommand \@@endlink[0]{}%
\providecommand \url  [0]{\begingroup\@sanitize@url \@url }%
\providecommand \@url [1]{\endgroup\@href {#1}{\urlprefix }}%
\providecommand \urlprefix  [0]{URL }%
\providecommand \Eprint [0]{\href }%
\providecommand \doibase [0]{http://dx.doi.org/}%
\providecommand \selectlanguage [0]{\@gobble}%
\providecommand \bibinfo  [0]{\@secondoftwo}%
\providecommand \bibfield  [0]{\@secondoftwo}%
\providecommand \translation [1]{[#1]}%
\providecommand \BibitemOpen [0]{}%
\providecommand \bibitemStop [0]{}%
\providecommand \bibitemNoStop [0]{.\EOS\space}%
\providecommand \EOS [0]{\spacefactor3000\relax}%
\providecommand \BibitemShut  [1]{\csname bibitem#1\endcsname}%
\let\auto@bib@innerbib\@empty
%</preamble>
\bibitem [{\citenamefont {Karsten}\ and\ \citenamefont
  {Smit}(1981)}]{karsten1981lattice}%
  \BibitemOpen
  \bibfield  {author} {\bibinfo {author} {\bibfnamefont {Luuk~H}\ \bibnamefont
  {Karsten}}\ and\ \bibinfo {author} {\bibfnamefont {Jan}\ \bibnamefont
  {Smit}},\ }\bibfield  {title} {\enquote {\bibinfo {title} {Lattice fermions:
  Species doubling, chiral invariance and the triangle anomaly},}\ }\href
  {\doibase 10.1016/0550-3213(81)90549-6} {\bibfield  {journal} {\bibinfo
  {journal} {Nucl. Phys. B}\ }\textbf {\bibinfo {volume} {183}},\ \bibinfo
  {pages} {103--140} (\bibinfo {year} {1981})}\BibitemShut {NoStop}%
\bibitem [{\citenamefont {Nielsen}\ and\ \citenamefont
  {Ninomiya}(1981)}]{Nielsen:1980rz}%
  \BibitemOpen
  \bibfield  {author} {\bibinfo {author} {\bibfnamefont {H.~B.}\ \bibnamefont
  {Nielsen}}\ and\ \bibinfo {author} {\bibfnamefont {M.}~\bibnamefont
  {Ninomiya}},\ }\bibfield  {title} {\enquote {\bibinfo {title} {{Absence of
  Neutrinos on a Lattice. 1. Proof by Homotopy Theory}},}\ }\href {\doibase
  10.1016/0550-3213(82)90011-6} {\bibfield  {journal} {\bibinfo  {journal}
  {Nucl. Phys. B}\ }\textbf {\bibinfo {volume} {185}},\ \bibinfo {pages} {20}
  (\bibinfo {year} {1981})},\ \bibinfo {note} {[Erratum: Nucl.Phys.B 195, 541
  (1982)]}\BibitemShut {NoStop}%
\bibitem [{\citenamefont {Kaplan}(2024)}]{kaplan2024chiral}%
  \BibitemOpen
  \bibfield  {author} {\bibinfo {author} {\bibfnamefont {David~B}\ \bibnamefont
  {Kaplan}},\ }\bibfield  {title} {\enquote {\bibinfo {title} {Chiral gauge
  theory at the boundary between topological phases},}\ }\href {\doibase
  10.1103/PhysRevLett.132.141603} {\bibfield  {journal} {\bibinfo  {journal}
  {Phys. Rev. Lett.}\ }\textbf {\bibinfo {volume} {132}},\ \bibinfo {pages}
  {141603} (\bibinfo {year} {2024})},\ \Eprint
  {http://arxiv.org/abs/2312.01494} {arXiv:2312.01494} \BibitemShut {NoStop}%
\bibitem [{\citenamefont {Clancy}\ and\ \citenamefont
  {Kaplan}(2025)}]{clancy2025chiral}%
  \BibitemOpen
  \bibfield  {author} {\bibinfo {author} {\bibfnamefont {Michael}\ \bibnamefont
  {Clancy}}\ and\ \bibinfo {author} {\bibfnamefont {David~B}\ \bibnamefont
  {Kaplan}},\ }\bibfield  {title} {\enquote {\bibinfo {title} {Chiral edge
  states on spheres for lattice domain wall fermions},}\ }\href {\doibase
  10.1103/PhysRevD.111.L031503} {\bibfield  {journal} {\bibinfo  {journal}
  {Physical Review D}\ }\textbf {\bibinfo {volume} {111}},\ \bibinfo {pages}
  {L031503} (\bibinfo {year} {2025})},\ \Eprint
  {http://arxiv.org/abs/2410.23065} {arXiv:2410.23065 [hep-lat]} \BibitemShut
  {NoStop}%
\bibitem [{\citenamefont {Kaplan}(1992)}]{Kaplan:1992bt}%
  \BibitemOpen
  \bibfield  {author} {\bibinfo {author} {\bibfnamefont {David~B.}\
  \bibnamefont {Kaplan}},\ }\bibfield  {title} {\enquote {\bibinfo {title} {{A
  method for simulating chiral fermions on the lattice}},}\ }\href {\doibase
  10.1016/0370-2693(92)91112-M} {\bibfield  {journal} {\bibinfo  {journal}
  {Phys. Lett. B}\ }\textbf {\bibinfo {volume} {288}},\ \bibinfo {pages}
  {342--347} (\bibinfo {year} {1992})},\ \Eprint
  {http://arxiv.org/abs/hep-lat/9206013} {arXiv:hep-lat/9206013} \BibitemShut
  {NoStop}%
\bibitem [{\citenamefont {Jansen}\ and\ \citenamefont
  {Schmaltz}(1992)}]{Jansen:1992tw}%
  \BibitemOpen
  \bibfield  {author} {\bibinfo {author} {\bibfnamefont {Karl}\ \bibnamefont
  {Jansen}}\ and\ \bibinfo {author} {\bibfnamefont {Martin}\ \bibnamefont
  {Schmaltz}},\ }\bibfield  {title} {\enquote {\bibinfo {title} {{Critical
  momenta of lattice chiral fermions}},}\ }\href {\doibase
  10.1016/0370-2693(92)91335-7} {\bibfield  {journal} {\bibinfo  {journal}
  {Phys. Lett. B}\ }\textbf {\bibinfo {volume} {296}},\ \bibinfo {pages}
  {374--378} (\bibinfo {year} {1992})},\ \Eprint
  {http://arxiv.org/abs/hep-lat/9209002} {arXiv:hep-lat/9209002} \BibitemShut
  {NoStop}%
\bibitem [{\citenamefont {Golterman}\ \emph {et~al.}(1993)\citenamefont
  {Golterman}, \citenamefont {Jansen},\ and\ \citenamefont
  {Kaplan}}]{Golterman:1992ub}%
  \BibitemOpen
  \bibfield  {author} {\bibinfo {author} {\bibfnamefont {Maarten F.~L.}\
  \bibnamefont {Golterman}}, \bibinfo {author} {\bibfnamefont {Karl}\
  \bibnamefont {Jansen}}, \ and\ \bibinfo {author} {\bibfnamefont {David~B.}\
  \bibnamefont {Kaplan}},\ }\bibfield  {title} {\enquote {\bibinfo {title}
  {Chern-{S}imons currents and chiral fermions on the lattice},}\ }\href
  {\doibase 10.1016/0370-2693(93)90692-B} {\bibfield  {journal} {\bibinfo
  {journal} {Phys. Lett. B}\ }\textbf {\bibinfo {volume} {301}},\ \bibinfo
  {pages} {219--223} (\bibinfo {year} {1993})},\ \Eprint
  {http://arxiv.org/abs/hep-lat/9209003} {arXiv:hep-lat/9209003} \BibitemShut
  {NoStop}%
\bibitem [{\citenamefont {Kaplan}\ and\ \citenamefont
  {Sen}(2024)}]{kaplan2024weyl}%
  \BibitemOpen
  \bibfield  {author} {\bibinfo {author} {\bibfnamefont {David~B}\ \bibnamefont
  {Kaplan}}\ and\ \bibinfo {author} {\bibfnamefont {Srimoyee}\ \bibnamefont
  {Sen}},\ }\bibfield  {title} {\enquote {\bibinfo {title} {Weyl fermions on a
  finite lattice},}\ }\href {\doibase 10.1103/PhysRevLett.132.141604}
  {\bibfield  {journal} {\bibinfo  {journal} {Phys. Rev. Lett.}\ }\textbf
  {\bibinfo {volume} {132}},\ \bibinfo {pages} {141604} (\bibinfo {year}
  {2024})},\ \Eprint {http://arxiv.org/abs/2312.04012} {arXiv:2312.04012}
  \BibitemShut {NoStop}%
\bibitem [{\citenamefont {Aoki}\ and\ \citenamefont
  {Fukaya}(2022)}]{aoki2022curved}%
  \BibitemOpen
  \bibfield  {author} {\bibinfo {author} {\bibfnamefont {Shoto}\ \bibnamefont
  {Aoki}}\ and\ \bibinfo {author} {\bibfnamefont {Hidenori}\ \bibnamefont
  {Fukaya}},\ }\bibfield  {title} {\enquote {\bibinfo {title} {Curved
  domain-wall fermions},}\ }\href {\doibase 10.1093/ptep/ptac075} {\bibfield
  {journal} {\bibinfo  {journal} {Progress of Theoretical and Experimental
  Physics}\ }\textbf {\bibinfo {volume} {2022}},\ \bibinfo {pages} {063B04}
  (\bibinfo {year} {2022})},\ \Eprint {http://arxiv.org/abs/2203.03782}
  {arXiv:2203.03782} \BibitemShut {NoStop}%
\bibitem [{\citenamefont {Aoki}\ and\ \citenamefont
  {Fukaya}(2023)}]{aoki2023curved}%
  \BibitemOpen
  \bibfield  {author} {\bibinfo {author} {\bibfnamefont {Shoto}\ \bibnamefont
  {Aoki}}\ and\ \bibinfo {author} {\bibfnamefont {Hidenori}\ \bibnamefont
  {Fukaya}},\ }\bibfield  {title} {\enquote {\bibinfo {title} {Curved
  domain-wall fermion and its anomaly inflow},}\ }\href {\doibase
  doi.org/10.1093/ptep/ptad023} {\bibfield  {journal} {\bibinfo  {journal}
  {Progress of Theoretical and Experimental Physics}\ }\textbf {\bibinfo
  {volume} {2023}},\ \bibinfo {pages} {033B05} (\bibinfo {year} {2023})},\
  \Eprint {http://arxiv.org/abs/2212.11583} {arXiv:2212.11583} \BibitemShut
  {NoStop}%
\bibitem [{\citenamefont {Aoki}\ \emph {et~al.}(2024)\citenamefont {Aoki},
  \citenamefont {Fukaya},\ and\ \citenamefont {Kan}}]{aoki2024}%
  \BibitemOpen
  \bibfield  {author} {\bibinfo {author} {\bibfnamefont {Shoto}\ \bibnamefont
  {Aoki}}, \bibinfo {author} {\bibfnamefont {Hidenori}\ \bibnamefont {Fukaya}},
  \ and\ \bibinfo {author} {\bibfnamefont {Naoto}\ \bibnamefont {Kan}},\
  }\bibfield  {title} {\enquote {\bibinfo {title} {A lattice formulation of
  weyl fermions on a single curved surface},}\ }\href {\doibase
  10.1093/ptep/ptae041} {\bibfield  {journal} {\bibinfo  {journal} {Progress of
  Theoretical and Experimental Physics}\ }\textbf {\bibinfo {volume} {2024}},\
  \bibinfo {pages} {043B05} (\bibinfo {year} {2024})},\ \Eprint
  {http://arxiv.org/abs/2402.09774} {arXiv:2402.09774} \BibitemShut {NoStop}%
\bibitem [{\citenamefont {Aoki}(2023)}]{aoki2024study}%
  \BibitemOpen
  \bibfield  {author} {\bibinfo {author} {\bibfnamefont {Shoto}\ \bibnamefont
  {Aoki}},\ }\emph {\bibinfo {title} {Study of Curved Domain-wall Fermions on a
  Lattice}},\ \href@noop {} {Ph.D. thesis} (\bibinfo {year} {2023}),\ \Eprint
  {http://arxiv.org/abs/2404.01002} {arXiv:2404.01002 [hep-lat]} \BibitemShut
  {NoStop}%
\bibitem [{\citenamefont {Grabowska}\ and\ \citenamefont
  {Kaplan}(2016)}]{grabowska2016nonperturbative}%
  \BibitemOpen
  \bibfield  {author} {\bibinfo {author} {\bibfnamefont {Dorota~M}\
  \bibnamefont {Grabowska}}\ and\ \bibinfo {author} {\bibfnamefont {David~B}\
  \bibnamefont {Kaplan}},\ }\bibfield  {title} {\enquote {\bibinfo {title}
  {Nonperturbative regulator for chiral gauge theories?}}\ }\href {\doibase
  10.1103/PhysRevLett.116.211602} {\bibfield  {journal} {\bibinfo  {journal}
  {Phys. Rev. Lett.}\ }\textbf {\bibinfo {volume} {116}},\ \bibinfo {pages}
  {211602} (\bibinfo {year} {2016})}\BibitemShut {NoStop}%
\bibitem [{\citenamefont {Callan}\ and\ \citenamefont
  {Harvey}(1985)}]{Callan:1984sa}%
  \BibitemOpen
  \bibfield  {author} {\bibinfo {author} {\bibfnamefont {Curtis~G.}\
  \bibnamefont {Callan}, \bibfnamefont {Jr.}}\ and\ \bibinfo {author}
  {\bibfnamefont {Jeffrey~A.}\ \bibnamefont {Harvey}},\ }\bibfield  {title}
  {\enquote {\bibinfo {title} {{Anomalies and fermion zero modes on strings and
  domain walls}},}\ }\href {\doibase 10.1016/0550-3213(85)90489-4} {\bibfield
  {journal} {\bibinfo  {journal} {Nucl. Phys. B}\ }\textbf {\bibinfo {volume}
  {250}},\ \bibinfo {pages} {427--436} (\bibinfo {year} {1985})}\BibitemShut
  {NoStop}%
\bibitem [{\citenamefont {Witten}\ and\ \citenamefont
  {Yonekura}(2022)}]{witten_anomaly_2020}%
  \BibitemOpen
  \bibfield  {author} {\bibinfo {author} {\bibfnamefont {Edward}\ \bibnamefont
  {Witten}}\ and\ \bibinfo {author} {\bibfnamefont {Kazuya}\ \bibnamefont
  {Yonekura}},\ }\href {\doibase 10.1142/9789811231711_0014} {\enquote
  {\bibinfo {title} {Anomaly inflow and the eta-invariant},}\ } (\bibinfo
  {year} {2022}),\ \Eprint {http://arxiv.org/abs/1909.08775} {arXiv:1909.08775}
  \BibitemShut {NoStop}%
\bibitem [{\citenamefont {Jansen}(1992)}]{jansen1992chiral}%
  \BibitemOpen
  \bibfield  {author} {\bibinfo {author} {\bibfnamefont {Karl}\ \bibnamefont
  {Jansen}},\ }\bibfield  {title} {\enquote {\bibinfo {title} {Chiral fermions
  and anomalies on a finite lattice},}\ }\href {\doibase
  10.1016/0370-2693(92)91113-N} {\bibfield  {journal} {\bibinfo  {journal}
  {Physics Letters B}\ }\textbf {\bibinfo {volume} {288}},\ \bibinfo {pages}
  {348--354} (\bibinfo {year} {1992})},\ \Eprint
  {http://arxiv.org/abs/hep-lat/9206014} {arXiv:hep-lat/9206014} \BibitemShut
  {NoStop}%
\bibitem [{\citenamefont {Golterman}\ and\ \citenamefont
  {Shamir}(2024)}]{golterman2024conserved}%
  \BibitemOpen
  \bibfield  {author} {\bibinfo {author} {\bibfnamefont {Maarten}\ \bibnamefont
  {Golterman}}\ and\ \bibinfo {author} {\bibfnamefont {Yigal}\ \bibnamefont
  {Shamir}},\ }\bibfield  {title} {\enquote {\bibinfo {title} {Conserved
  currents in five-dimensional proposals for lattice chiral gauge theories},}\
  }\href {\doibase 10.1103/PhysRevD.109.114519} {\bibfield  {journal} {\bibinfo
   {journal} {Phys. Rev. D}\ }\textbf {\bibinfo {volume} {109}},\ \bibinfo
  {pages} {114519} (\bibinfo {year} {2024})},\ \Eprint
  {http://arxiv.org/abs/2404.16372} {arXiv:2404.16372} \BibitemShut {NoStop}%
\bibitem [{\citenamefont {Hamada}\ and\ \citenamefont
  {Kawai}(2017)}]{hamada2017axial}%
  \BibitemOpen
  \bibfield  {author} {\bibinfo {author} {\bibfnamefont {Yu}~\bibnamefont
  {Hamada}}\ and\ \bibinfo {author} {\bibfnamefont {Hikaru}\ \bibnamefont
  {Kawai}},\ }\bibfield  {title} {\enquote {\bibinfo {title} {{Axial U(1)
  current in Grabowska and Kaplan’s formulation}},}\ }\href {\doibase
  10.1093/ptep/ptx086} {\bibfield  {journal} {\bibinfo  {journal} {Progress of
  Theoretical and Experimental Physics}\ }\textbf {\bibinfo {volume} {2017}},\
  \bibinfo {pages} {063B09} (\bibinfo {year} {2017})},\ \Eprint
  {http://arxiv.org/abs/1705.01317} {arXiv:1705.01317} \BibitemShut {NoStop}%
\bibitem [{\citenamefont {Aoki}\ \emph {et~al.}(2023)\citenamefont {Aoki},
  \citenamefont {Fukaya}, \citenamefont {Kan}, \citenamefont {Koshino},\ and\
  \citenamefont {Matsuki}}]{aoki2023magnetic}%
  \BibitemOpen
  \bibfield  {author} {\bibinfo {author} {\bibfnamefont {Shoto}\ \bibnamefont
  {Aoki}}, \bibinfo {author} {\bibfnamefont {Hidenori}\ \bibnamefont {Fukaya}},
  \bibinfo {author} {\bibfnamefont {Naoto}\ \bibnamefont {Kan}}, \bibinfo
  {author} {\bibfnamefont {Mikito}\ \bibnamefont {Koshino}}, \ and\ \bibinfo
  {author} {\bibfnamefont {Yoshiyuki}\ \bibnamefont {Matsuki}},\ }\bibfield
  {title} {\enquote {\bibinfo {title} {Magnetic monopole becomes dyon in
  topological insulators},}\ }\href {\doibase 10.1103/PhysRevB.108.155104}
  {\bibfield  {journal} {\bibinfo  {journal} {Physical Review B}\ }\textbf
  {\bibinfo {volume} {108}},\ \bibinfo {pages} {155104} (\bibinfo {year}
  {2023})},\ \Eprint {http://arxiv.org/abs/2304.13954} {arXiv:2304.13954}
  \BibitemShut {NoStop}%
\bibitem [{\citenamefont {'t~Hooft}(1976)}]{t1976symmetry}%
  \BibitemOpen
  \bibfield  {author} {\bibinfo {author} {\bibfnamefont {Gerard}\ \bibnamefont
  {'t~Hooft}},\ }\bibfield  {title} {\enquote {\bibinfo {title} {Symmetry
  breaking through {B}ell-{J}ackiw anomalies},}\ }\href {\doibase
  10.1103/PhysRevLett.37.8} {\bibfield  {journal} {\bibinfo  {journal}
  {Physical Review Letters}\ }\textbf {\bibinfo {volume} {37}},\ \bibinfo
  {pages} {8--11} (\bibinfo {year} {1976})}\BibitemShut {NoStop}%
\bibitem [{\citenamefont {Banks}\ \emph {et~al.}(1994)\citenamefont {Banks},
  \citenamefont {Nir},\ and\ \citenamefont {Seiberg}}]{banks1994missing}%
  \BibitemOpen
  \bibfield  {author} {\bibinfo {author} {\bibfnamefont {Tom}\ \bibnamefont
  {Banks}}, \bibinfo {author} {\bibfnamefont {Yosef}\ \bibnamefont {Nir}}, \
  and\ \bibinfo {author} {\bibfnamefont {Nathan}\ \bibnamefont {Seiberg}},\
  }\bibfield  {title} {\enquote {\bibinfo {title} {Missing (up) mass,
  accidental anomalous symmetries, and the strong {CP} problem},}\ }\href@noop
  {} {\bibfield  {journal} {\bibinfo  {journal} {arXiv preprint
  hep-ph/9403203}\ } (\bibinfo {year} {1994})}\BibitemShut {NoStop}%
\bibitem [{\citenamefont {Perez}\ \emph {et~al.}(1994)\citenamefont {Perez},
  \citenamefont {Gonzalez-Arroyo}, \citenamefont {Snippe},\ and\ \citenamefont
  {van Baal}}]{perez1994instantons}%
  \BibitemOpen
  \bibfield  {author} {\bibinfo {author} {\bibfnamefont {Margarita~Garcia}\
  \bibnamefont {Perez}}, \bibinfo {author} {\bibfnamefont {Antonio}\
  \bibnamefont {Gonzalez-Arroyo}}, \bibinfo {author} {\bibfnamefont {Jeroen}\
  \bibnamefont {Snippe}}, \ and\ \bibinfo {author} {\bibfnamefont {Pierre}\
  \bibnamefont {van Baal}},\ }\bibfield  {title} {\enquote {\bibinfo {title}
  {Instantons from over-improved cooling},}\ }\href {\doibase
  https://doi.org/10.1016/0550-3213(94)90631-9} {\bibfield  {journal} {\bibinfo
   {journal} {Nuclear Physics B}\ }\textbf {\bibinfo {volume} {413}},\ \bibinfo
  {pages} {535--552} (\bibinfo {year} {1994})},\ \Eprint
  {http://arxiv.org/abs/hep-lat/9701012} {arXiv:hep-lat/9701012 [hep-lat]}
  \BibitemShut {NoStop}%
\bibitem [{\citenamefont {De~Forcrand}\ \emph {et~al.}(1997)\citenamefont
  {De~Forcrand}, \citenamefont {Perez},\ and\ \citenamefont
  {Stamatescu}}]{de1997topology}%
  \BibitemOpen
  \bibfield  {author} {\bibinfo {author} {\bibfnamefont {Philippe}\
  \bibnamefont {De~Forcrand}}, \bibinfo {author} {\bibfnamefont
  {Margarita~Garcia}\ \bibnamefont {Perez}}, \ and\ \bibinfo {author}
  {\bibfnamefont {Ion-Olimpiu}\ \bibnamefont {Stamatescu}},\ }\bibfield
  {title} {\enquote {\bibinfo {title} {Topology of the {SU}(2) vacuum: a
  lattice study using improved cooling},}\ }\href {\doibase
  https://doi.org/10.1016/S0550-3213(97)00275-7} {\bibfield  {journal}
  {\bibinfo  {journal} {Nuclear Physics B}\ }\textbf {\bibinfo {volume}
  {499}},\ \bibinfo {pages} {409--449} (\bibinfo {year} {1997})},\ \Eprint
  {http://arxiv.org/abs/hep-lat/9701012} {arXiv:hep-lat/9701012 [hep-lat]}
  \BibitemShut {NoStop}%
\bibitem [{\citenamefont {Banks}\ \emph {et~al.}(1991)\citenamefont {Banks},
  \citenamefont {Dine},\ and\ \citenamefont {Seiberg}}]{banks1991irrational}%
  \BibitemOpen
  \bibfield  {author} {\bibinfo {author} {\bibfnamefont {Tom}\ \bibnamefont
  {Banks}}, \bibinfo {author} {\bibfnamefont {Michael}\ \bibnamefont {Dine}}, \
  and\ \bibinfo {author} {\bibfnamefont {Nathan}\ \bibnamefont {Seiberg}},\
  }\bibfield  {title} {\enquote {\bibinfo {title} {Irrational axions as a
  solution of the strong {CP} problem in an eternal universe},}\ }\href
  {\doibase 10.1016/0370-2693(91)90561-4} {\bibfield  {journal} {\bibinfo
  {journal} {Physics Letters B}\ }\textbf {\bibinfo {volume} {273}},\ \bibinfo
  {pages} {105--110} (\bibinfo {year} {1991})}\BibitemShut {NoStop}%
\bibitem [{\citenamefont {Leutwyler}\ and\ \citenamefont
  {Smilga}(1992)}]{leutwyler1992spectrum}%
  \BibitemOpen
  \bibfield  {author} {\bibinfo {author} {\bibfnamefont {H}~\bibnamefont
  {Leutwyler}}\ and\ \bibinfo {author} {\bibfnamefont {A}~\bibnamefont
  {Smilga}},\ }\bibfield  {title} {\enquote {\bibinfo {title} {Spectrum of
  {D}irac operator and role of winding number in {QCD}},}\ }\href {\doibase
  10.1103/PhysRevD.46.5607} {\bibfield  {journal} {\bibinfo  {journal}
  {Physical Review D}\ }\textbf {\bibinfo {volume} {46}},\ \bibinfo {pages}
  {5607} (\bibinfo {year} {1992})}\BibitemShut {NoStop}%
\bibitem [{\citenamefont {Aoki}\ \emph {et~al.}(2025)\citenamefont {Aoki},
  \citenamefont {Blum}, \citenamefont {Collins}, \citenamefont {Debbio},
  \citenamefont {Morte}, \citenamefont {Dimopoulos}, \citenamefont {Feng},
  \citenamefont {Golterman}, \citenamefont {Gottlieb}, \citenamefont {Gupta},
  \citenamefont {Herdoiza}, \citenamefont {Hernandez}, \citenamefont
  {Jüttner}, \citenamefont {Kaneko}, \citenamefont {Lunghi}, \citenamefont
  {Meinel}, \citenamefont {Monahan}, \citenamefont {Nicholson}, \citenamefont
  {Onogi}, \citenamefont {Petreczky}, \citenamefont {Portelli}, \citenamefont
  {Ramos}, \citenamefont {Sharpe}, \citenamefont {Simone}, \citenamefont
  {Sint}, \citenamefont {Sommer}, \citenamefont {Tantalo}, \citenamefont
  {de~Water}, \citenamefont {Vaquero}, \citenamefont {Wenger},\ and\
  \citenamefont {Wittig}}]{aoki2025flagreview2024}%
  \BibitemOpen
  \bibfield  {author} {\bibinfo {author} {\bibfnamefont {Y.}~\bibnamefont
  {Aoki}}, \bibinfo {author} {\bibfnamefont {T.}~\bibnamefont {Blum}}, \bibinfo
  {author} {\bibfnamefont {S.}~\bibnamefont {Collins}}, \bibinfo {author}
  {\bibfnamefont {L.~Del}\ \bibnamefont {Debbio}}, \bibinfo {author}
  {\bibfnamefont {M.~Della}\ \bibnamefont {Morte}}, \bibinfo {author}
  {\bibfnamefont {P.}~\bibnamefont {Dimopoulos}}, \bibinfo {author}
  {\bibfnamefont {X.}~\bibnamefont {Feng}}, \bibinfo {author} {\bibfnamefont
  {M.}~\bibnamefont {Golterman}}, \bibinfo {author} {\bibfnamefont {Steven}\
  \bibnamefont {Gottlieb}}, \bibinfo {author} {\bibfnamefont {R.}~\bibnamefont
  {Gupta}}, \bibinfo {author} {\bibfnamefont {G.}~\bibnamefont {Herdoiza}},
  \bibinfo {author} {\bibfnamefont {P.}~\bibnamefont {Hernandez}}, \bibinfo
  {author} {\bibfnamefont {A.}~\bibnamefont {Jüttner}}, \bibinfo {author}
  {\bibfnamefont {T.}~\bibnamefont {Kaneko}}, \bibinfo {author} {\bibfnamefont
  {E.}~\bibnamefont {Lunghi}}, \bibinfo {author} {\bibfnamefont
  {S.}~\bibnamefont {Meinel}}, \bibinfo {author} {\bibfnamefont
  {C.}~\bibnamefont {Monahan}}, \bibinfo {author} {\bibfnamefont
  {A.}~\bibnamefont {Nicholson}}, \bibinfo {author} {\bibfnamefont
  {T.}~\bibnamefont {Onogi}}, \bibinfo {author} {\bibfnamefont
  {P.}~\bibnamefont {Petreczky}}, \bibinfo {author} {\bibfnamefont
  {A.}~\bibnamefont {Portelli}}, \bibinfo {author} {\bibfnamefont
  {A.}~\bibnamefont {Ramos}}, \bibinfo {author} {\bibfnamefont {S.~R.}\
  \bibnamefont {Sharpe}}, \bibinfo {author} {\bibfnamefont {J.~N.}\
  \bibnamefont {Simone}}, \bibinfo {author} {\bibfnamefont {S.}~\bibnamefont
  {Sint}}, \bibinfo {author} {\bibfnamefont {R.}~\bibnamefont {Sommer}},
  \bibinfo {author} {\bibfnamefont {N.}~\bibnamefont {Tantalo}}, \bibinfo
  {author} {\bibfnamefont {R.~Van}\ \bibnamefont {de~Water}}, \bibinfo {author}
  {\bibfnamefont {A.}~\bibnamefont {Vaquero}}, \bibinfo {author} {\bibfnamefont
  {U.}~\bibnamefont {Wenger}}, \ and\ \bibinfo {author} {\bibfnamefont
  {H.}~\bibnamefont {Wittig}},\ }\href {https://arxiv.org/abs/2411.04268}
  {\enquote {\bibinfo {title} {Flag review 2024},}\ } (\bibinfo {year}
  {2025}),\ \Eprint {http://arxiv.org/abs/2411.04268} {arXiv:2411.04268
  [hep-lat]} \BibitemShut {NoStop}%
\end{thebibliography}%
